\documentclass[aps,prd,12point,twocolumn,nofootinbib,showpacs,superscriptaddress]{revtex4-1}
\def\theequation{\arabic{section}.\arabic{equation}}
\usepackage{psfrag}
\usepackage{subfigure}
\usepackage{color}
\usepackage{mathrsfs}
\usepackage{graphicx}
\usepackage{amssymb, bm}
\usepackage{amsmath, amsthm}
\usepackage{epstopdf}
\usepackage{hyperref}
\usepackage{enumerate}
\usepackage{longtable}

\newcommand{\be}{\begin{equation}}
\newcommand{\ee}{\end{equation}}

\begin{document}
\def\theequation{\arabic{section}.\arabic{equation}}


\title{Analogy between equilibrium beach profiles and closed universes}


\author{Valerio Faraoni}
\email[]{vfaraoni@ubishops.ca}
\affiliation{Department of Physics \& Astronomy, Bishop's University\\
2600 College Street, Sherbrooke, Qu\'ebec, Canada J1M~1Z7
}


\begin{abstract}

We reformulate the variational problem describing equilibrium beach 
profiles in the thermodynamic approach of Jenkins and Inman. A first 
integral of the resulting Euler-Lagrange equation coincides formally with 
the Friedmann equation ruling closed universes in relativistic cosmology, 
leading to a useful analogy. Using the machinery of 
Friedmann-Lema\^itre-Robertson-Walker cosmology, qualitative properties 
and analytic solutions of beach profiles, which are the subject of a 
controversy, are elucidated.

\end{abstract}


\maketitle

\section{Introduction}
\label{sec:1}
\setcounter{equation}{0}

Since the early work of Bruun \cite{Bruun}, the profile of a beach, 
measured from the shore seaward and perpendicular to 
the shoreline, has been one of the most studied features of coastal 
morphology. It  is important not only from the scientific point of view, 
but also 
because of its relevance to human activities \cite{profiles-general} 
(early research was motivated by interest in military operations). A 
beach profile is 
dynamical and undergoes 
seasonal changes \cite{changes}, therefore research has 
focussed on the 
simpler problem of  {\em equilibrium} beach profiles, on which there is a 
significant literature  \cite{profiles-general, JI06}. Data show an 
undulating relief where the landward 
side of the topography increases for a while, while the seaward side 
decreases \cite{DolanDean85, Otvos00, RuessinkTerwindt00}. A beach 
profile is then modelled by matching two 
different curves, each of which satisfies an appropriate ordinary 
differential equation ({\em e.g.}, \cite{JI06}).

Research on the subject has moved from mere data-fitting to developing  
theories of beach profiles under different conditions ({\em e.g.}, 
breaking or non-breaking waves). The most promising approach is probably 
that of 
Jenkins and Inman \cite{JI06}, which is based on thermodynamics. Near the 
shore, wave motion causes turbulence 
and energy dissipation and the main idea of Ref.~\cite{JI06} consists of    
maximizing the rate of energy dissipation of  both breaking and 
non-breaking waves.  This extremization leads to an elegant variational 
principle formulation of the problem and to an associated Euler-Lagrange 
equation for the curves describing the equilibrium beach profiles. Since 
this equation is non-linear, the search for its solutions 
is non-trivial. Analytic solutions were proposed in \cite{JI06}, 
but they are not easily reproducible and have recently been criticized in 
\cite{MaldonadoUchasara}.  

Instead of formulating the variational problem for a functional of the 
beach profile $h(x)$, Ref.~\cite{JI06} expresses it in terms of the 
inverse function $x(h)$. We reformulate the problem in terms of $h(x)$ and 
it is then easy to find a first integral of the Euler-Lagrange equation 
arising from a symmetry. The key point of the present work is the 
realization that this first integral is formally equivalent to the 
Friedmann equation  ruling the evolution of closed universes in 
relativistic cosmology, provided that the cosmic fluid that causes their 
spacetime curvature is of a specific type. This fluid is indeed very 
reasonable from the physical point of view. The cosmological analogy turns 
out to be very useful because a wealth of information is now available 
about  the equations of relativistic cosmology and their solutions. 
Research in cosmology has been much more intensive, and dates back to the 
1920s (see, {\em e.g.}, \cite{BookofUniverses} for a historical 
perspective), which is longer than the time spanned by the research on 
beach profiles. We apply the 
standard exact solutions of the Einstein-Friedmann 
equations of cosmology \cite{Wald, Carroll, Liddle, KT}, supplemented 
by recent mathematical results and methods for the Friedmann 
equation \cite{Chen0, Chen15b, Chen15a}, to the analog 
beach problem. This use of the analogy leads us to clarifying several 
issues about beach profiles and to a 
comprehensive treatment of analytic solutions of  
the non-linear differential equation ruling beach profiles in the 
thermodynamic approach of \cite{JI06}. 

While it is understandable that the cosmological analogy was missed in 
the literature because of the enormous gap between the communities of 
cosmologists and ocean scientists, it is surprising that another, rather  
obvious, analogy between any beach profile ODE and the one-dimensional 
motion of  a point particle was also missed. While this second analogy is 
much less useful than the first one, it nevertheless  provides some 
insight on the qualitative nature of the solutions of the beach 
profile  equation, and we discuss it briefly.

The structure of this paper is as follows: In Sec.~\ref{sec:2} we 
reformulate the Jenkins-Inman variational problem and we rewrite the 
resulting first integral of (our version of) the Euler-Lagrange equation 
in a form analogous to the Friedmann equation. Section~\ref{sec:3} 
discusses 
the mechanical 
analogy. Section~\ref{sec:4} develops the cosmological analogy, while 
Sec.~\ref{sec:5} discusses in detail the analytic solutions of the 
beach profile  equation and their deep water approximation. 
Section~\ref{sec:6} contains a 
summary and the conclusions.  We follow the notation of Ref.~\cite{Wald}; 
the signature of the spacetime metric is $-+++$, and we use units in which 
Newton's constant $G$ and the speed of light $c$ are unity.

\section{Equilibrium beach profiles}
\label{sec:2}
\setcounter{equation}{0}

\begin{figure}
\includegraphics[scale=0.55]{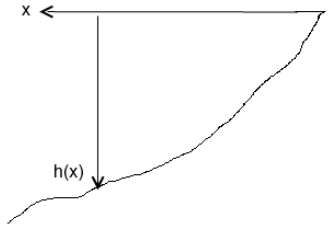}
\caption{The $x$-axis points seaward horizontally from the 
shore and $h(x)$, measured downward, is the local 
water depth.\label{fig:0}} 
\end{figure}

Let $x$ be the cross-shore distance (the $x$-axis is horizontal and 
pointing seaward) and $h(x) $ be the local water depth, measured downward 
from a  (constant) mean sea level (Fig.~\ref{fig:0}). 
The authors of \cite{JI06} seek to maximize the entropy by  extremizing  
the functional 
\be
I \left[ x(h) \right] = \int_{h_1}^{h_2} \left( 
h(x)\right)^{\frac{-\,3(n+1)}{4} } 
\sqrt{ 1+\left( 
\frac{dx}{dh} \right)^2} dh \,,
\ee
where $n>0$ is an exponent appearing in the relation between the shear 
stress amplitude $\tau_0$ and the water velocity $u_m(x)$ at the sea floor
\be
\tau_0(x) = K_{\tau} \rho u_m^n(x) \,.
\ee
Here $\rho$ is the seawater density and the proportionality 
constant $K_{\tau}$ is independent of $u_m$ \cite{JI06}. 

Instead of studying the variational principle $\delta I=0$ for $x(h)$, it 
is convenient to recast the problem in terms of the actual depth 
profile $h(x)$ as\footnote{For $n=-7/3$, the Lagrangian reduces to 
$L=h\sqrt{ 1+ \left( dh/dx \right)^2 } $ and gives rise to the 
classic catenary problem \cite{Goldstein, Boas}, but negative values of 
$n$ are excluded in \cite{JI06}.} 
\be
J\left[ h(x) \right] =\int_{x_1}^{x_2} dx \left( h(x) \right)^{\frac{ -3\, 
(n+1)}{4}} \sqrt{ 
1+ \left( \frac{dh}{dx} \right)^2 } \,. \label{eq:3} 
\ee
The Lagrangian is
\be
L\left( h, h'\right)= \left( h(x) \right)^{\frac{ -3\, (n+1)}{4}} \sqrt{ 
1+ \left( h' \right)^2 }  \,, \label{eq:4}
\ee
where $h'\equiv dh/dx$. Since $\partial L/\partial x=0$, the Hamiltonian
\be
{\cal H}=  p_h h'-L(h,h') 
\ee 
is conserved, where
\be
p_h \equiv \frac{\partial L}{\partial h'} 
=\frac{ h^{\frac{ -3\, (n+1)}{4} } h'}{  \sqrt{ 
1+ \left( h' \right)^2 }} \label{eq:6}
\ee
is the momentum canonically conjugated to $h$. The conservation of 
\be
{\cal H}= - \frac{1}{   h^{\frac{ 3\, (n+1)}{4}}\,    \sqrt{ 
1+  h'^2 } }\label{eq:7}
\ee
yields the first integral of motion
\be
h^{\frac{ 3\, (n+1)}{2}} \left( 1+  h'^2 \right)=C^2 \,,\label{eq:8}
\ee
where $C$ is an integration constant. It is clear that it must be $C\neq 
0$, otherwise the solution is $h(x)=0$ everywhere.  Imposing the boundary 
condition of zero depth at the origin, $h(0)=0$, rules out any constant 
solutions (which would be unphysical anyway) and forces $h'(x)$ to diverge 
as $x\rightarrow 0$ in order to keep the left hand side of 
Eq.~(\ref{eq:8}) constant.  The presence of this cusp prevents the 
applicability of the usual existence and uniqueness theorems for   
the initial value problem at $x=0$ \cite{BrauerNoel}.\footnote{Curiously, 
this situation resembles the fact that the longitudinal profile of a 
glacier as described by the Vialov equation of glaciology necessarily has  
a cusp at its terminus. This is because the Vialov ODE exhibits a 
feature similar to  
Eq.~(\ref{eq:8}) \cite{Paterson,Hooke, GreveBlatter, Hutter, glaxshape}.} 
A physical consquence of this cusp is that the shallow water approximation 
used in \cite{JI06} breaks down near the shore.

Equation~(\ref{eq:8}) can be re-arranged as 
\be
\left( \frac{h'}{h} \right)^2 = \frac{C^2}{h^{\frac{3n+7}{2} } }   
-\frac{1}{h^2} \,. \label{eq:9}
\ee
This equation is formally the same as the Friedmann equation ruling the 
evolution of certain spatially homogeneous and isotropic  
(Friedmann-Lema\^itre-Robertson-Walker, in short ``FLRW'') universes in   
general relativity \cite{Wald, Carroll, Liddle, KT}. This fact gives rise 
to a 
very useful formal analogy between equilibrium beach profiles and closed 
universes in Einstein's theory of gravity. Given that the study of the 
cosmological equations has a long history \cite{BookofUniverses}, it is 
easy to infer 
mathematical solutions for the analog beach profile problem. Moreover, 
recent results on the mathematical properties of solutions of the 
Friedmann equation play a significant role. As we shall see, the analogy 
sheds some light on 
the mathematical solutions of Eq.~(\ref{eq:8}) describing beach profiles, 
which are currently the subject of a 
controversy \cite{MaldonadoUchasara}. Note that the analogy with cosmology 
emerges only when the variational problem for the beach profiles is 
formulated in terms of $h(x)$ instead of $x(h)$. Before discussing it, 
however, it is useful to visit another analogy (missed in the 
literature thus far) between equilibrium beach 
profiles and point particle mechanics, which illustrates 
graphically certain 
qualitative properties of the solutions of Eq.~(\ref{eq:8}).

\section{Mechanical analogy}
\label{sec:3}
\setcounter{equation}{0}

\begin{figure}
\includegraphics[scale=0.35]{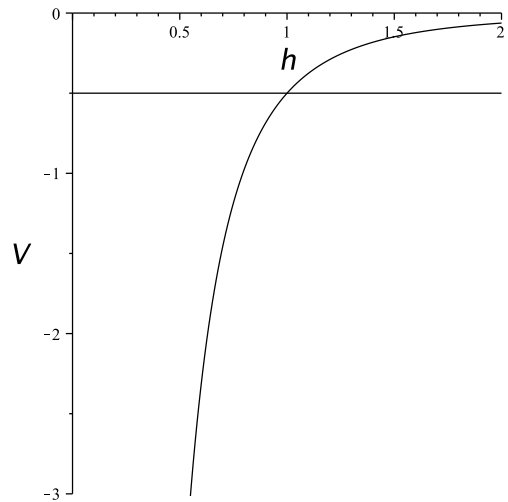}%
\caption{In the region $h>0$ there is always a unique intersection 
between the horizontal line $E=-1/2$ and the potential energy $V(h)$, 
therefore the motion is always confined between the origin and the turning 
point.\label{fig:1}}
\end{figure}

Let us rewrite the ordinary differential equation~(\ref{eq:8}) as 
\be
\frac{h'^2}{2} + V(h)=E \,,\label{m1}
\ee
where
\be
V(h)=-\frac{C^2}{2 h^{3(n+1)/2}} \label{m2}
\ee
and $E=-1/2$. In the form~(\ref{m1}), Eq.~(\ref{eq:8}) can be interpreted 
formally as describing as the position of a particle of unit mass and 
kinetic energy $(h')^2/2$ 
in one-dimensional motion along 
the $h$-axis, subject to the potential energy $V(h)$, as time $x$ goes by. 
Since this 
fictitious  particle is subject only to the conservative force $-dV/dh$, 
its total mechanical energy is conserved and has the constant value  
$E=-1/2$. Equation~(\ref{m1}) is a first integral of Newton's second law 
$d^2 h/dx^2 =-dV/dh$ expressing energy conservation. Following the 
Weierstrass approach 
 \cite{Goldstein, BochicchioLaserra07, Destradeetal07, 
Bochicchioetal11}, one obtains a qualitative 
understanding and a graphical representation of the possible motions 
({\em 
i.e.}, of the 
possible  solutions of Eq.~(\ref{eq:8}))  from the graph of the potential 
$V(h)$  and its intersections with the  horizontal line $E = - 1/2$ (see 
Fig.~\ref{fig:1}).
 
The function $V(h)$ has a vertical asymptote at $h=0$, the $h$-axis 
as a horizontal asymptote, and only the region 
$h \geq 0$ is physical.  Since $E\geq V$, the possible motions ({\em 
i.e.}, the solutions $h(x)$ of Eq.~(\ref{eq:8}))  are always 
confined to the interval $0\leq h\leq h_*$, where the turning 
point $h_*$ is the horizontal coordinate of the unique intersection 
between the line $E=-1/2$ and $V(h)$. This turning point is 
\be
\left( h_*, V_* \right)= \left( \left( \frac{C^2}{2|E|} \right)^{ 
\frac{2}{3(n+1)} },  E  \right)  
\ee
and is always present for all negative energies $E$, in 
particular for $E=-1/2$. It is unique.  
A solution $h(x)$ of Eq.~(\ref{eq:8}) describes only a 
segment of the beach profile \cite{DolanDean85, Otvos00, 
RuessinkTerwindt00} (in Ref.~\cite{JI06}, two ellipsoidal 
cycloids are matched at a point to form a realistic profile).

Since $V(h)\rightarrow -\infty$ as $h\rightarrow 0^{+}$, a particle 
approaching $h=0$ from the right must have diverging kinetic energy to 
keep the total energy $E$ finite (and equal to $-1/2$). This means that 
it is always $h' \rightarrow + \infty$ as $h\rightarrow 0^+$ (which we had 
already concluded 
by inspection of Eq.~(\ref{eq:8})). The origin of this divergence can be 
traced to the fact that Eq.~(\ref{eq:8}) was derived in \cite{JI06} 
under 
the approximation of a mild slope of the profile
\be
\frac{ \tan \beta}{kh}=\frac{h'}{kh} \ll 1 \,,
\ee
where $k$ is the wave vector and $\tan \beta =h'$ is the bottom 
slope. It is shown in Ref.~\cite{JI06} that $k 
\simeq \frac{\omega}{ \sqrt{gh}}$ (where $\omega$ is the angular frequeny 
of the breaking wave and $g$ is the acceleration of 
gravity), which yields
\be
\frac{h'}{kh} \simeq \frac{h'}{\sqrt{h}} \rightarrow \infty \;\;\;\;\; 
\mbox{as} \;\;\; x,h\rightarrow 0^{+} \,,
\ee
violating the mild slope approximation near the shore. There is nothing 
else to gain from the analogy with the one-dimensional motion of a point 
particle, and we now turn to the richer analogy with cosmology.

\section{Cosmological analogy}
\label{sec:4}
\setcounter{equation}{0}

Here we recall the essentials of FLRW cosmology and develop the analogy 
with equilibrium beach profiles. 

In relativistic cosmology, the geometry of a  spatially homogeneous and 
isotropic universe is necessarily given by the four-dimensional FLRW line 
element
\be
ds^2 = -dt^2 +a^2(t) \left[ \frac{dr^2}{1-Kr^2} +r^2 \left( d\theta^2 + 
\sin^2 \theta \, d\varphi^2 \right)\right] \,, \label{eq:10}
\ee
written here in comoving polar coordinates $\left(t, r, \theta, \varphi 
\right)$. The scale factor $a(t)$ describes how two points at 
fixed comoving  coordinate distance $r_0$ (for example, two typical 
galaxies without proper motions) separate as the universe expands. At 
time $t$, the 
physical distance between these two points is $ l=a(t)r_0$ and it 
increases 
if $a(t)$ increases to describe an expanding universe. The function $a(t)$ 
embodies the expansion history of the universe. The constant $K$ in 
Eq.~(\ref{eq:10}) is normalized to the only three possible values $K=1, 0, 
-1$ describing, respectively, a closed universe (closed 3-dimensional 
spatial sections $t=$~const.), Euclidean spatial sections, or hyperbolic 
3-spaces \cite{Wald, Carroll, Liddle, KT}. This classification includes 
all the possible FLRW geometries 
and all the dynamics is encoded in the evolution of the scale factor 
$a(t)$ as a function of the comoving time $t$.

It is common in cosmology to describe the matter content of the universe, 
which generates the spacetime curvature, as a perfect fluid of energy 
density $\rho(t)$ and isotropic pressure $P(t)$ related by some equation 
of state. The functions $a(t), \rho(t)$, and $P(t)$ satisfy the 
Einstein-Friedmann equations
\begin{eqnarray}
&&H^2 \equiv \left( \frac{\dot{a}}{a}\right)^2 =\frac{8\pi}{3} \, \rho 
-\frac{K}{a^2} \,, \label{eq:11}\\
&&\nonumber\\
&&\frac{\ddot{a}}{a}= -\, \frac{4\pi}{3} \left( \rho +3P \right) \,, 
\label{eq:12} \\
&&\nonumber\\
&& \dot{\rho}+3H\left(P+\rho \right)=0 \,,\label{eq:13}
\end{eqnarray}
where an overdot denotes differentiation with respect to $t$ and 
$H(t)\equiv 
\dot{a}/a$ is the Hubble parameter \cite{Wald, Carroll, Liddle, KT}. Only 
two 
of 
these three equations are independent; given any two, the third one can be 
derived from them. For convenience, and without loss of 
generality, we take the Friedmann equation~(\ref{eq:11}) and the energy 
conservation equation~(\ref{eq:13}) as primary, and the acceleration 
equation~(\ref{eq:12}) as derived. 

Equation~(\ref{eq:11}) with $K=+1$ is formally the same as 
Eq.~(\ref{eq:9}) ruling equilibrium beach profiles if we exchange the 
variables $\left( x, h(x) \right) \longrightarrow \left( t, a(t) \right)$. 
The analogy holds if a suitable cosmological fluid fills the analog 
universe. By comparing Eqs.~(\ref{eq:11}) and~(\ref{eq:9}), we see that it 
must be
\be
\rho(t)=\frac{\rho_0}{ \left( a(t) \right)^{ \frac{3n+7}{2} }} 
\,,\label{eq:14} 
\ee 
where $\rho_0 $ is a positive integration constant determined by the 
initial conditions. This relation is familiar in cosmology, where it is 
common to assume that the cosmic fluid satisfies the barotropic equation 
of 
state 
\be
P=w\rho \label{eq:15}
\ee
for a suitable constant $w$ (``equation of state 
parameter'').\footnote{The 
assumption that $w$ is constant is often relaxed \cite{Liddle, KT}, but 
this 
complication is not necessary, nor useful, here.} Then Eq.~(\ref{eq:13}) 
is integrated to give
\be
\rho(a) = \frac{ \rho_0}{ a^{3(w+1)} }  \label{eq:16} \,.
\ee
By comparing Eqs.~(\ref{eq:14}) and~(\ref{eq:16}), one concludes that the 
analogy between beach profiles and cosmology is valid if the universe is 
filled with a perfect  fluid with $P=w\rho$ and equation of state 
parameter 
\be
w=\frac{3n+1}{6} \,. \label{eq:17}
\ee
Since it must be $n>0$ in the model of Ref.~\cite{JI06}, it is $w>1/6$. 
Well known cases discussed in cosmology textbooks are a radiation fluid 
$w=1/3$ (corresponding to $n=1/3$) and a stiff fluid $w=1$ (corresponding 
to $n=5/3$), which is realized by a  free scalar field acting as an 
effective fluid \cite{Wald, Carroll, Liddle, KT}. 

Since $w>1/6$, the acceleration equation~(\ref{eq:12}) implies that the 
analog universe always decelerates, {\em i.e.}, $\ddot{a}<0$ (only if 
$P<-\rho/3$ does the universe accelerate, as is clear by inspecting the 
right hand side of the acceleration equation~(\ref{eq:12})).

\section{Solutions of the beach profile equation via Friedmann analogue}
\label{sec:5}
\setcounter{equation}{0}

Let us analyze the solutions of Eq.~(\ref{eq:9}), which are the subject of 
an ongoing controversy \cite{MaldonadoUchasara}, in the light of the 
analog Friedmann equation. It 
is convenient to begin with the simplest case (we refer the reader to 
standard textbooks ({\em e.g.}, \cite{Goldstein, Boas}) for the classic 
catenary problem obtained for the unphysical value $n=-7/3$).

\subsection{The case $n=1/3$}

In the special case $n=1/3$, corresponding to $w=1/3$ in the analog 
universe dominated by a gas of photons, Eq.~(\ref{eq:14}) gives the 
typical blackbody scaling of the energy density $\rho(a)=\rho_0/a^4$ and 
the scale factor \cite{Wald, Carroll, Liddle, KT}
\be
a(t) = \sqrt{C'} \sqrt{ 1-\left( 1- \frac{t}{\sqrt{C'}} \right)^2} 
\,,\label{eq:18}
\ee
where $C'$ is a positive integration constant. This solution describes 
a closed universe that begins at a Big Bang singularity $a=0$ at $t=0$, 
expands to a maximum size $\sqrt{C'}$, and collapses to a Big Crunch 
singularity at $t=2\sqrt{C'}$.  The 
corresponding equilibrium beach profile is 
\be
h(x)=h_0 \sqrt{ 1-\left( 1- \frac{x}{h_0} \right)^2} 
\,,\label{eq:18bis}
\ee
with $h_0$ a constant length. The graph of $h(x) $ in the interval $x\in 
\left( 0, 2h_0\right)$ is a cycloid (a semi-circle), {\em i.e.}, the 
trajectory of a 
point located on the rim of a circle of radius $h_0$ that rolls without 
slipping on the $x$-axis.

\subsection{The value $n=-1$ (linearly expanding universe)}

Other special cases give simple exact solutions well known in cosmology, 
but they correspond to negative values of $n$, which are unphysical in the 
thermodynamic model of \cite{JI06}. We report them here nevertheless. 

If $w=-1/3$, corresponding to $n=-1$, the acceleration 
equation~(\ref{eq:12}) gives the linear solution. In terms of the analog 
beach profile, it is
\be
h(x)=h_0x+h_1 \,. 
\ee
Linear beach profiles are considered in~\cite{MaldonadoUchasara} and, in 
the shallow water approximation, they are reported in \cite{Dean91}. 

It is easy to see that a linear solution is the only possible power law 
solution of Eq.~(\ref{eq:8}) (here we refer to {\em exact} solutions: 
approximate solutions can be power law, as we will see later). In fact, 
assuming $h(x)=A x^{\alpha}$ with 
$A$ and $\alpha$ constants, substitution into Eq.~(\ref{eq:8}) yields 
immediately $\left( n, \alpha \right)=  \left( -1, 1 \right)$  and 
$1+A=\pm C$.

\subsection{The value $n=-1/3$ (cosmic dust)}

Another special case corresponds to a cosmic dust fluid $w=0$, obtained 
for $n=-1/3$. In this case the explicit solution in parametric form is 
\cite{Wald, Carroll, Liddle, KT}
\begin{eqnarray}
h(\eta) &=& \frac{C}{2}\left( 1-\cos\eta \right) \,,\\
&&\nonumber\\
x(\eta) &=& \frac{C}{2}\left( \eta -\sin\eta \right) \,.
\end{eqnarray}
Expanding for $\eta \ll 1$ yields
\begin{eqnarray}
h(\eta) & \simeq & \frac{C}{4} \, \eta^2 \,,\\
&&\nonumber\\
x(\eta) & \simeq & \frac{C}{12}\, \eta^3 \,.
\end{eqnarray}
Then, $h/x \approx 3/\eta \gg 1$, which shows the meaning of the 
approximation 
$\eta \ll1$: it corresponds to deep water. By eliminating the parameter 
$\eta$, one obtains
\be
h(x) \simeq \left( \frac{9C}{4}\right)^{1/3} x^{2/3} \,.
\ee
This profile was obtained in Ref.~\cite{Dean91} and claimed to be a good 
fit to field data.

\subsection{The general case $w=$~const.}

In the general case $w=$~const., a solution of the cosmological 
equations~(\ref{eq:11})-(\ref{eq:13}) can be found in parametric 
form and up to a quadrature by performing a change of variable 
\cite{Landau}. Let us 
adopt the conformal time $\eta$ defined by $dt=ad\eta$. Then the 
Einstein-Friedmann equations give
\be
\eta  =\pm \int \frac{da}{a \sqrt{\frac{8\pi }{3} \, \rho a^2 -K}} 
\,.\label{eq:19}
\ee
When $w=$~const., the substitution of Eq.~(\ref{eq:16}) yields
\be
\eta  =\pm \int \frac{da}{a \sqrt{\frac{8\pi }{3} \, a^{-(3w+1)}  -K}} 
\,.\label{eq:20}
\ee
By introducing the rescaled variable
\be \label{eq:21}
z \equiv \left( \frac{8\pi  C_1}{3} \right)^{\frac{-1}{3w+1}} a
\ee
and using, for $K=+1$, 
\be
\int \frac{dz}{z\sqrt{ z^m-1}}=\frac{2}{m} \mbox{arcsec} \left( 
z^{m/2}\right) 
\,,
\ee
one integrates Eq.~(\ref{eq:20}) and inverts the result, obtaining the 
parametric solution with conformal time as the parameter \cite{Landau, 
AmJP1, Chen0} 
\begin{eqnarray}
a(\eta) &=& a_0 \left[ \cos\left( c\eta+d \right) \right]^{1/c} 
\,,\label{eq:23} \\
&&\nonumber\\
t(\eta) &=& a_0 \int_0^{\eta} d\eta' \left[ \cos\left( c\eta'+d \right) 
\right]^{1/c}  \,, \label{eq:24}
\end{eqnarray}
where 
\be
c=\frac{3w+1}{2} \label{eq:25}
\ee
and $a_0$ is a constant. The Big Bang boundary condition $a=0$ at $t=0$ 
(corresponding to $\eta=0$)  is satisfied if $d=-\pi/2$, which yields
\begin{eqnarray}
a(\eta) &=& a_0 \left[ \sin \left( \frac{(3w+1}{2} \, \eta \right) 
\right]^{\frac{2}{3w+1}} 
\,,\label{eq:23} \\
&&\nonumber\\
t(\eta) &=& a_0 \int_0^{\eta} d\eta' \left[ \sin\left( 
\frac{(3w+1}{2} \, \eta'  \right) \right]^{\frac{2}{3w+1}}  \,. 
\label{eq:24}
\end{eqnarray}

On the beach profile side, the analog of the conformal time parameter is 
defined by $d\eta=dx/h(x)$. Small increments of the dimensionless 
parameter $\eta$ are small increments of the distance from the 
shoreline measured in units of the local water depth. In finite terms, 
Eq.~(\ref{eq:20}) has the analogue
\be
\eta=\pm \int \frac{dh}{h\sqrt{ \frac{8\pi}{3} \, h^{\frac{-(3n+7)}{2} } 
-1}} \,,\label{eq:26}
\ee
which integrates to
\begin{eqnarray}
h(\eta) &=& h_0 \left[ \sin\left( \frac{3(n+1)}{4} \, \eta  \right) 
\right]^{\frac{4}{3(n+1)} } \,,\label{eq:27} \\
&&\nonumber\\
x(\eta) &=& h_0 \int_0^{\eta} d\eta' \left[ \sin\left( 
\frac{3(n+1)}{4} \, \eta'\right) 
\right]^{\frac{4}{3(n+1)} }  \,, \label{eq:28}
\end{eqnarray}
where $x(\eta)$ is reduced to a quadrature and $c$ is given by 
Eq.~(\ref{eq:25}). 

In the special case $n=1/3$ considered in the previous subsection it is 
$c=1$, the integration of Eq.~(\ref{eq:28}) is trivial, and the parameter 
$\eta$ can be eliminated obtaining the explicit solution $h(x)$ given by 
Eq.~(\ref{eq:18}). 

An alternative way to solve for the cosmic dynamics consists of reasoning 
on the acceleration equation and noting that, in conformal time $\eta$, 
the 
latter reduces to a Riccati equation \cite{AmJP1}. Assuming that 
$w=$~const., the acceleration equation~(\ref{eq:12}) becomes 
\be
\frac{\ddot{a}}{a} +c \, \frac{\dot{a}^2}{a^2} +\frac{cK}{a^2}=0 \,.
\ee
For $K=+1$ and using conformal time, this equation is re-written as
\be
\frac{1}{a} \, \frac{d^2a}{d\eta^2} +\frac{ (c-1)}{a^2} \left( 
\frac{da}{d\eta} \right)^2 +c=0 \,.\label{Riccati}
\ee
This standard Riccati equation \cite{Ince, Hille} is solved by using the 
new variable 
\be
u \equiv \frac{1}{a} \, \frac{da}{d\eta} 
\ee
and then setting 
\be
u \equiv \frac{1}{cv}\, \frac{dv}{d\eta} \,,
\ee
which reduces the Riccati equation~(\ref{Riccati}) to the harmonic 
oscillator equation $v''+c^2 v=0$, with sine and cosine solutions. Going 
back to the original variable $a (\eta)$ reproduces the 
solution~(\ref{eq:27}) and (\ref{eq:28}) \cite{AmJP1}.

It is, of course, interesting to know when the solution can be expressed 
explicitly in terms of elementary functions, as in the case $n=1/3 $ 
discussed above. This question is answered in 
Ref.~\cite{Chen0} with the help of the Chebysev theorem of integration 
\cite{Chebysev, MarchisottoZakeri}. Manipulation of the Friedmann 
equation~(\ref{eq:11}) yields \cite{Chen0}
\be
t= \int da \, \frac{ a^{ \frac{3w+1}{2} } }{\sqrt{\frac{8\pi\rho_0}{3} 
-a^{3w+1} } } 
\ee
or, introducing \cite{Chen0}
\be
b_0 \equiv \frac{8\pi \rho_0}{3} \,, \;\;\;\;\;\; u \equiv a^{ 
\frac{3(w+1)}{2} } \,,
\ee
it is
\be
t=\frac{2}{3(w+1)} \int \frac{du}{\sqrt{ b_0-u^\gamma}} \,,
\ee
where 
\be
\gamma=\frac{2(3w+1)}{3(w+1)} 
\ee
for $ w\neq -1$. According to Chebysev's theorem, the integral is 
elementary only if  $1/\gamma$ or $\frac{2-\gamma}{2\gamma}$ is an integer 
\cite{Chen0}.
Setting $1/\gamma = N=0, \pm 1, \pm 2\, \pm 3 , \, ...$ yields 
$w=\frac{3-2N}{3(2N-1)} $ and 
\be
n=\frac{7-6N}{3(2N-1)} \,.
\ee
The requirement of Ref.~\cite{JI06} that $n>0$ corresponds to 
$\frac{1}{2}< N<\frac{7}{6}$, which leaves only $N=1$, corresponding to 
$n=w=1/3$. The other possibility $\frac{2-\gamma}{2\gamma}=N $ corresponds 
to  $w=\frac{1-N}{3N} $ and to $n=(2-3N)/(3N)$. The requirement $n>0$ is 
then 
equivalent to $ 0<N<2/3$, which is not satisfied by any integer.

\subsection{Deep water approximation}

We can now derive a deep water approximation for the general 
solution~(\ref{eq:27}) and~(\ref{eq:28}). Expanding these equations for 
$\eta \ll 1$ yields
\begin{eqnarray}
h(\eta) & \simeq & h_0 \left( \frac{3(n+1)}{4}\right)^{ \frac{4}{3(n+1)}} 
\eta^{ \frac{4}{3(n+1)}} \,,\\
&&\nonumber\\
x(\eta) & \simeq & h_0 \left( \frac{3(n+1)}{4}\right)^{ \frac{4}{3(n+1)}} 
\frac{3(n+1)}{7+3n }  \, 
\eta^{ \frac{7+3n}{3(n+1)}  }  \,.
\end{eqnarray} 
We have 
\be
\frac{h}{x} \simeq \frac{(7+3n)}{3(n+1)}  
\, \frac{1}{\eta} \gg 1 
\ee
independent of the value of $n$. Therefore, $\eta \ll 1$ corresponds 
to the deep water approximation.  Eliminating the parameter $\eta$, we 
obtain the approximate power law solution
\be
h(x) \simeq h_0^{\frac{3(n+1)}{7+3n}} \left( \frac{7+3n}{4} \right)^{ 
\frac{4}{7+3n}}  \, x^{\frac{4}{7+3n} } 
\ee
(power law beach profiles have been proposed since the early studies of 
this subject \cite{Bruun, Bowen80, Dean91}). The power is equal to $2/3$ 
(the value 
advocated in Ref.~\cite{Dean91}) 
if $n=-1/3$, as already seen in a special case. The different exponent 
$2/5$ advocated in 
\cite{Bowen80} is achieved for $n=1$. For all other values of $n$, the 
exponent is instead $4/(7+3n)$.

\subsection{Roulettes}

The qualitative study and the search for analytic solutions of the 
Einstein-Friedmann equations~(\ref{eq:11})-(\ref{eq:13}) are reviewed in 
\cite{oldAmJP, AmJP1, SonegoTalamini}), while \cite{Chen0, Chen15a, 
Chen15b} 
report new efforts in this direction. A mathematical property of the 
Friedmann equation~(\ref{eq:11}) 
demonstrated in~\cite{Chen15b} is that the graphs of all 
solutions of this equation are roulettes. A roulette is the locus of 
a point that lies on, or inside, a curve that rolls without slipping on a  
straight line.\footnote{In a more general definition, the curve rolls 
without 
slipping along another curve, but this is an  unnecessary 
complication here.} Indeed, all the solutions of the beach profile 
equation~(\ref{eq:9}) proposed in \cite{JI06} have graphs that are 
elliptical cycloids, {\em i.e.}, the curves described by a point on an 
ellipse as the latter rolls on the $x$-axis. In the special case in which 
the ellipse reduces to a circle, one obtains an ordinary cycloid (a 
semi-circle like the one given by Eq.~(\ref{eq:18bis})). Chen {\em et al.} 
\cite{Chen15b} study explicitly the Friedmann equation for a closed 
($K=+1$) universe to derive the equation of the solution in polar 
coordinates $\left(r , \vartheta \right)$. We do not repeat their 
analysis, reporting only the results. In general, $ r(\vartheta)$ is not 
explicit and is only obtained up to a quadrature, 
but there are integrable cases corresponding to particular fluids 
with energy density
\be
\rho(a) = \frac{\alpha}{a^2} +\beta a^{\delta} \,,
\ee
where $\alpha, \beta$, and $\delta$ are arbitrary constants 
\cite{Chen15b} (although 
it must be $\alpha \geq 0$ and $\beta \geq 0$ to avoid 
negative densities). Our case is reproduced for $\alpha=0, \beta=\rho_0$, 
and 
$\delta=-(3n+7)/2$. The solution, constructed as a roulette, is 
\cite{Chen15b}
\be
\frac{1}{ r^{ \frac{3n+7}{3n+1}} } =\cos\left( \frac{3n+7}{3n+1} 
\,\vartheta \right) \,. \label{roulette}
\ee
Particularly simple solutions correspond to $w=0$ ($n=-1/3$, discussed 
separately in \cite{Chen15b}) and $w=1/9$ 
($n=-1/9$) which are excluded in the model of Ref.~\cite{JI06}. As already 
mentioned, all the solutions proposed in \cite{JI06} are roulettes, but 
they are not reproduced by Eq.~(\ref{roulette}) (see also 
Ref.~\cite{MaldonadoUchasara}).

\section{Summary and Conclusions}
\label{sec:6}
\setcounter{equation}{0}

Using an analogy with relativistic cosmology and, to a much 
lesser extent, a  different one with one-dimensional point 
particle motion, we have derived and studied the non-linear 
ODE~(\ref{eq:8})  ruling beach profiles in the Jenkins-Inman thermodynamic 
approach to  the problem of equilibrium beach profiles \cite{JI06}.  
Contrary to  Ref.~\cite{JI06}, we 
first reformulate the variational principle in terms of the beach profile 
$h(x)$, instead of its inverse $x(h)$,\footnote{Our equation~(\ref{eq:8}) 
is not contained in Ref.~\cite{JI06}, 
although the resulting beach profiles should be the same as those 
obtainable by these authors' equations once their function $x(h)$ is 
inverted.} which uncovers two  analogies.  

The first is an analogy with the mechanics of a point particle in 
one-dimensional motion, which provides a graphic way of deducing basic 
qualitative properties of the solutions. The second, and much richer, 
analogy is with relativistic cosmology, as described by Einstein's theory 
of general relativity. It is rather surprising that there is a formal 
analogy between the Friedmann equation describing closed universes and the 
beach profile equation. Since there are {\em two} independent 
equations ruling the evolution of these universes, one extra condition 
must be imposed, {\em i.e.}, the cosmic perfect fluid sourcing the analog 
universe must have a specific equation of state. {\em A priori}, this 
extra condition would be expected to generate a completly exotic fluid 
with an 
unphysical equation of state, which would make the analogy far less 
interesting. A similar analogy for the transversal ({\em i.e.}, 
cross-sectional) 
profile of glaciated valleys holds \cite{facets}: in that case, the cosmic 
fluid is very exotic, with a non-linear equation of state, albeit of a 
type considerd by cosmologists studying dark energy \cite{nonlinear-eos, 
Chen15a}. 
In the beach profile analogy, however, the cosmic fluid required is  
physically very reasonable: its equation of state is 
barotropic, linear, and constant. This type of cosmic fluid is very 
common in the cosmology textbooks \cite{Wald, Carroll, Liddle, KT} and 
includes, as a special case, a  
radiation fluid ({\em i.e.}, an expanding blackbody distribution of 
incoherent photons with random phases, polarizations, and directions of 
propagation) describing the radiation era of the early universe 
\cite{Carroll, Wald, Liddle, KT}.

Since there is  a wealth of literature on analytic solutions of the 
Friedmann equation, one can use the analogy beach profiles-closed 
universes to discover the solutions of the beach profile  
equation~(\ref{eq:8}), which are currently the subject of a controversy 
\cite{MaldonadoUchasara}. The solutions can be given in parametric 
form $\left( h(\eta), x(\eta) \right)$ with $x(\eta)$ expressed up to a 
quadrature. The Jenkins-Inman formalism contains  another parameter $n$ 
which is related 
to shear stress and water velocity at the sea floor and is also related 
to the equation of state parameter of the cosmic fluid in the analogous 
universe.  Special  values of this parameter $n$ corresponding to 
integrability of the first order ODE~(\ref{eq:8}) have been identified, 
and some simple exact solutions provided.  Furthermore, recent results 
\cite{Chen15b} demonstrate that 
all the solutions of the Friedmann equation and, therefore, all those of 
the beach profile equation~(\ref{eq:8}), are roulettes. The solutions 
proposed (in polar coordinates) in Ref.~\cite{JI06}) are indeed roulettes, 
but their form is not reproduced by the integrability cases listed in 
\cite{Chen15b}, lending support to the critique of 
\cite{MaldonadoUchasara}. At the end of the day, however, much is learned 
about analytic solutions for beach profiles in the thermodynamic 
approach thanks to the cosmological analogy (and, to a much lesser extent, 
to the mechanical analogy) developed here. Three-dimensional  beach 
profiles not contemplated in \cite{JI06} would be analogous to anisotropic 
universes (Bianchi models) in relativistic cosmology \cite{KSMcH}, and 
will be studied in  the future.


\begin{acknowledgments}

This work is supported, in part,  by Bishop's University and by the 
Natural Sciences \& Engineering Research Council of Canada (Grant No.
2016-03803). 
\end{acknowledgments}

\end{document}